\documentclass[aps,prl,twocolumn,superscriptaddress,showpacs,amsmath,amssymb]{revtex4}

\usepackage[dvips]{graphicx}
\usepackage{bm}
\usepackage{ulem}

\newcommand{\be}{\begin{equation}}
\newcommand{\ee}{\end{equation}}
\newcommand{\grad}{\bm\nabla}
\newcommand{\curl}{\bm\nabla\wedge}
\newcommand{\divr}{\bm\nabla\cdot}

\begin{document}

\title{Breathing modes of a fast rotating Fermi gas}
\author{Mauro Antezza}
\affiliation{Dipartimento di Fisica, Universit\`a di Trento and CNR-INFM R\&D
Center on Bose-Einstein Condensation, Via Sommarive 14, I-38050 Povo, Trento,
Italy}
\author{Marco Cozzini}
\affiliation{Dipartimento di Fisica, Politecnico di Torino, Corso Duca degli
Abruzzi 24, I-10129 Torino, Italy}
\affiliation{Quantum Information Group, Institute for Scientific Interchange
(ISI), Viale Settimio Severo 65, I-10133 Torino, Italy}
\author{Sandro Stringari}
\affiliation{Dipartimento di Fisica, Universit\`a di Trento and CNR-INFM R\&D
Center on Bose-Einstein Condensation, Via Sommarive 14, I-38050 Povo, Trento,
Italy}


\begin{abstract}
We derive the frequency spectrum of the lowest compressional 
oscillations of a 3D harmonically trapped Fermi superfluid in the presence of a
vortex lattice, treated in the diffused vorticity approximation within a hydrodynamic approach. We consider the
general case of a superfluid at $T=0$ characterized by a polytropic equation of
state ($\sim{n}^\gamma$), which includes both the Bose-Einstein condensed regime
of dimers ($\gamma=1$) and the unitary limit of infinite scattering length
($\gamma=2/3$). Important
limiting cases are considered, including the centrifugal limit, the isotropic
trapping and  the cigar geometry. The conditions required to enter the lowest Landau level and quantum Hall regimes at unitarity are also discussed. 
\end{abstract}
\maketitle
The experimental realization of quantized vortices in interacting ultracold Fermi
gases \cite{Zwierlein05} has opened new challenging perspectives in the
experimental and theoretical study of superfluidity. 
These perspectives are particularly important because in Fermi gases the 
measurement of the order parameter is not directly accessible. Actually, only for small and positive values of the scattering
length, when dimers built up with pairs of atoms of opposite spin are formed, the
fermionic system gives rise to the phenomenon of Bose-Einstein condensation
(BEC), whose onset is clearly revealed by the bimodal structure of the density \cite{BECmolecules}. The
observation of vortices for negative values of the scattering length as well as
in the  unitary limit close to a Feshbach resonance, where the scattering
length is larger than the interparticle distance, consequently provides a unique
source of information on the superfluid nature of these novel configurations. 

The measurements of Ref.~\cite{Zwierlein05} have however shown that vortices
in atomic Fermi gases are not directly observable {\it in situ} nor
after expansion, unless one suddenly ramps the scattering length to small and
positive values (corresponding to the BEC regime) just after the release of the trap.
Indeed, the visibility of vortices is limited in both the BCS and unitary limits, in the first case due to the reduced contrast and in the
latter mainly due to the smallness of their size, fixed by the 
interparticle distance.

For the above reasons it is interesting to explore more macroscopic signatures of
the presence of vortices. A first important source of information comes from
the bulge effect associated with the increase of the radial size of the cloud
produced by the centrifugal force. More systematic information comes from the
study of the collective oscillations. For example the splitting of the quadrupole
frequencies with opposite angular momentum \cite{zambelli} provides direct
information on the angular momentum carried by the vortical configuration and has
been used to measure even the quantization of a single vortex line  in
Bose-Einstein condensed atomic gases \cite{dalibard}. In this work we focus on
the study of the compressional modes whose frequency is affected by the
presence of the vortex lines and, at the same time, is sensitive to the nature of
the  configuration (BEC gas of dimers, unitary limit, etc.). 
The study of the compressional modes actually provides a unique information on
the  equation of state of these systems. In the absence of rotation it has been
the object of recent theoretical \cite{Stringari04,grigory} and experimental
\cite{thomas,grimm1,grimm2} work.   
Theoretical studies of the collective oscillations in rotating  Fermi gases  
were so far limited to small angular velocities in the absence of vortices
\cite{Cozzini03Fermi}. First calculations based on the diffused vorticity
approximation were recently carried out in cylindrical geometry \cite{Machida06}. 

We consider here a two-component Fermi gas with balanced spin population,  trapped by a three-dimensional
axisymmetric harmonic potential, so that the collective oscillations can be
labeled by the axial component $m$ of angular momentum. 
The typical wavelength of the lowest  modes is of the order of the system size. When
the number of vortices in the sample is large, this length scale is 
much larger than the intervortex distance. In order to evaluate the corresponding
oscillation frequencies one can then rely on a coarse grain description of the
system, the so-called diffused vorticity approximation
\cite{Feynman55,Cozzini03BosoniRot}, which does not require to deal with the
microscopic details of single vortices. By considering the case of a regular
lattice of singly quantized vortices, this long-wavelength description assumes
the vorticity to be uniformly spread in the fluid. In practice, if the vortex
lattice rotates at angular velocity $\bm\Omega$, the average curl of the velocity
field $\bm{v}$ is given by $\curl\bm{v}=2\bm\Omega$, characterizing the rigid
body rotation $\bm{v}=\bm\Omega\wedge\bm{r}$.
This also corresponds to a uniform vortex density, which, for a Fermi superfluid,
is ${n}_{\text{v}}=2M\Omega/\pi\hbar$, i.e., a factor 2 larger than in the case
of a Bose superfluid with the same value of the atomic mass $M$. 

The main consequence of
the diffused vorticity approximation is  the introduction of  an effective velocity field which does not
satisfy anymore the irrotationality constraint of the microscopic superfluid
flow, but accounts for  the presence of the vortex lattice.
Within this framework, we consider the   problem of solving the equations of 
rotational hydrodynamics with a polytropic equation of state, where the
chemical potential $\mu$ is assumed to have a power law dependence on the density
${n}$, namely, $\mu\propto{n}^\gamma$. This parametrization treats exactly  several important configurations of interacting Fermi gases, including the  Bose-Einstein
condensed regime of dimers ($\gamma=1$), where the scattering length is small and positive, and the unitary regime of infinite scattering length where the equation of state takes a universal density dependence characterized by the value 
$\gamma=2/3$. Moreover, as discussed in the literature (see, e.g.,
Ref.~\cite{grigory} and references therein), it is possible to provide an accurate, although approximate, description of  the entire BEC-BCS crossover by introducing an effective exponent $\gamma$ for the
equation of state. Since
the compressional modes are  sensitive to the equation of state,
the accurate study of their frequency can then provide a useful insight on the various regimes achieved in the experiments.

The equations of rotational hydrodynamics, written, in the laboratory frame, are given by
\begin{eqnarray}
\!\!\!\!\!\!\partial_t{n}+\divr({n}\bm{v}) \!& = & \!0 \ ,
\label{Continuity}\\
\!\!\!\!\!\!M\partial_t\bm{v}+\grad\left(Mv^2/2+
V_{\text{ext}}+\mu_{\text{loc}}\right)\! & = &\!
M\bm{v}\wedge(\curl\bm{v}) \ , \label{Euler}
\end{eqnarray}
where $\mu_{\text{loc}}(\bm{r},t)\propto{n}^\gamma(\bm{r},t)$ is the local
chemical potential fixed by the equation of state of the  uniform matter,
$\bm{v}(\bm{r},t)$ is the velocity field, and
$V_{\text{ext}}$ is the external potential which is assumed to be the same for
both the spin components of the Fermi gas. For an axisymmetric harmonic
potential  $V_{\text{ext}}=M\left[\omega_{\perp}^2(x^2+y^2)+\omega_z^2
z^2\right]/2$ and for a rotation of the trap in the $x$-$y$ plane at frequency
$\Omega_0$, the equilibrium solutions of Eqs.~(\ref{Continuity}) and
(\ref{Euler}) are given by $\bm{v}_0(\bm{r})  =  \bm\Omega_0\wedge\bm{r}$ and 
${n}_0(\bm{r})  \propto 
[\mu_0-\tilde{V}_{\text{ext}}(\bm{r})]^{1/\gamma}$, where
$\tilde{V}_{\text{ext}}= V_{\text{ext}}- M \Omega_0^2(x^2+y^2)/2
$
is the renormalized trapping potential accounting for the centrifugal effect
produced by the rotation and the chemical potential $\mu_0$ is obtained from the
normalization condition for the density. The centrifugal force causes a bulge effect which modifies the aspect ratio of the rotating cloud
according to the relationship 
%
\begin{equation}
R^2_z/ R^2_\perp= \left(\omega^2_\perp-\Omega_0^2\right)/ \omega^2_z,
\label{bulge}
\end{equation}
where 
$R_z$ and $R_\perp$ are, respectively, the axial and radial Thomas-Fermi radii of the cloud.
It also fixes a natural limit for the angular velocity $\Omega_0$ which cannot exceed the radial trapping frequency $\omega_\perp$.

By expanding Eqs.~(\ref{Continuity}) and (\ref{Euler}) with respect to small
perturbations of the density and velocity field, ${n}={n}_0+\delta{n}$ and
$\bm{v}=\bm{v}_0+\delta\bm{v}$, one obtains two coupled linearized
equations
which admit several solutions of relevant physical interest.
On the one side, one has surface solutions carrying angular momentum and
characterized by irrotational flow. For example, the most relevant $m=\pm2$
quadrupole solutions are described by density variations of the form
$\delta{n}\propto(x\pm iy)^2$ and by the velocity field
$\delta\bm{v}\propto\nabla(x\pm iy)^2$
\cite{Cozzini03BosoniRot,Machida06,Bigelow03,Chevy03}.
These surface modes exhibit the dispersion $\omega_\pm  =
\sqrt{2\omega^2_\perp-\Omega_0^2}\pm \Omega_0$ and are  strongly
affected by the rotational effect, as experimentally proven in the
case of Bose-Einstein condensed atomic gases \cite{jilam=2}. Their frequency is
however independent of the equation of state and is not expected to exhibit a
new behavior in the case of a Fermi superfluid.

In addition to the surface modes the hydrodynamic equations exhibit an important
class of $m=0$ compressional  modes. In order to solve the linearized equations
of motion we use the Ansatz \cite{Cozzini03BosoniRot,Cozzini03Fermi}
\begin{eqnarray}
\delta\bm{v} &=& \left\{\delta\bm\Omega \wedge \bm{r}+ \nabla \left[\alpha_{\perp} (x^2+y^2)+\alpha_z z^2\right]\right\}
e^{-i\omega t} \ , \label{vANS}\\
\delta{n}&=&{n}_0^{1-\gamma}\left[a_0+a_{\perp}(x^2+y^2)+a_zz^2\right]
e^{-i\omega t} \ , \label{nANS}
\end{eqnarray}
where $\delta\bm\Omega$, parallel to the axial direction, accounts for the proper
variation of the angular velocity during the oscillation. The inclusion of this
term is crucial to ensure the conservation of angular momentum. It was ignored in
previous  works \cite{Machida06,Bigelow03} on the collective oscillations of
two-dimensional Bose and Fermi gases containing a vortex lattice where pure
\textit{irrotational} flow  was assumed.  While the irrotational assumption is
valid for the $m\neq0$ modes, it turns out to be inadequate for the compressional
$m=0$ oscillations which are associated with variations of the vortex density and
hence of the angular velocity.

The Ansatz (\ref{vANS}-\ref{nANS}) gives rise to a linear system yielding three
solutions for the oscillation frequency $\omega$.
One of them is the trivial solution $\omega=0$, corresponding to a
change of the equilibrium configuration due to the adiabatic change of the
angular velocity of the system.
The other two solutions instead correspond to the radial and axial breathing modes of the gas and their frequency is given by
\begin{widetext}
\begin{eqnarray}
\lefteqn{\!\!\!\!\!\!\!\!
\omega_\pm^2 = (1+\gamma)\omega_{\perp}^2
+\frac{2+\gamma}{2}\,\omega_z^2+(1-\gamma)\Omega_0^2} \nonumber \\
&&\!\!\!\!\!\!\!\!\!\!\!\!\pm\sqrt{(1+\gamma)^2\omega_{\perp}^4
+\left(\frac{2+\gamma}{2}\right)^2\omega_z^4+(1-\gamma)^2\Omega_0^4
+(\gamma^2-3\gamma-2)\omega_z^2\omega_{\perp}^2
+2(1+\gamma)(1-\gamma)\omega_{\perp}^2\Omega_0^2
-(\gamma^2-\gamma+2)\omega_z^2\Omega_0^2} \ . 
\label{finalresult}
\end{eqnarray}
\end{widetext}
Equation (\ref{finalresult}) represents the main result of the present work.
These two solutions arise from  the coupling between the radial and axial motion
caused by the hydrodynamic forces in the presence of the rigid rotation of the
gas. It is easy to see that Eq.~(\ref{finalresult}) reproduces, as limiting
cases, the result of Ref.~\cite{Stringari96} for the non-rotating Bose condensed
gas ($\Omega_0=0$, $\gamma=1$), the result of Ref.~\cite{Cozzini03BosoniRot} for
the rotational Bose gas ($\Omega_0\neq0$, $\gamma=1$), and finally the result of 
Ref.~\cite{Cozzini03Fermi} for the non-rotating superfluid Fermi gas
($\Omega_0=0$, for generic $\gamma$).

\begin{figure}
\begin{center}
\includegraphics[width=0.45\textwidth]{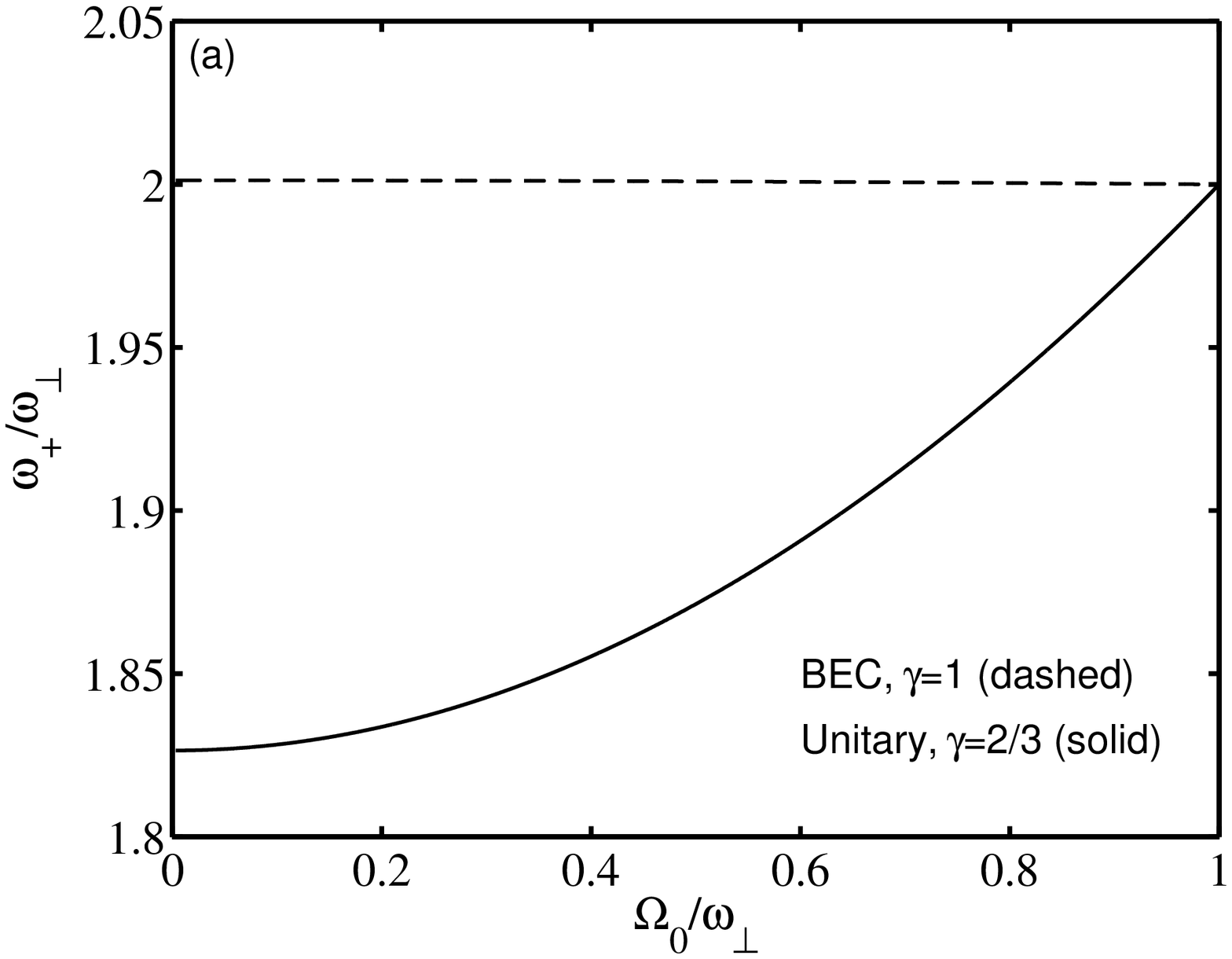}
\includegraphics[width=0.45\textwidth]{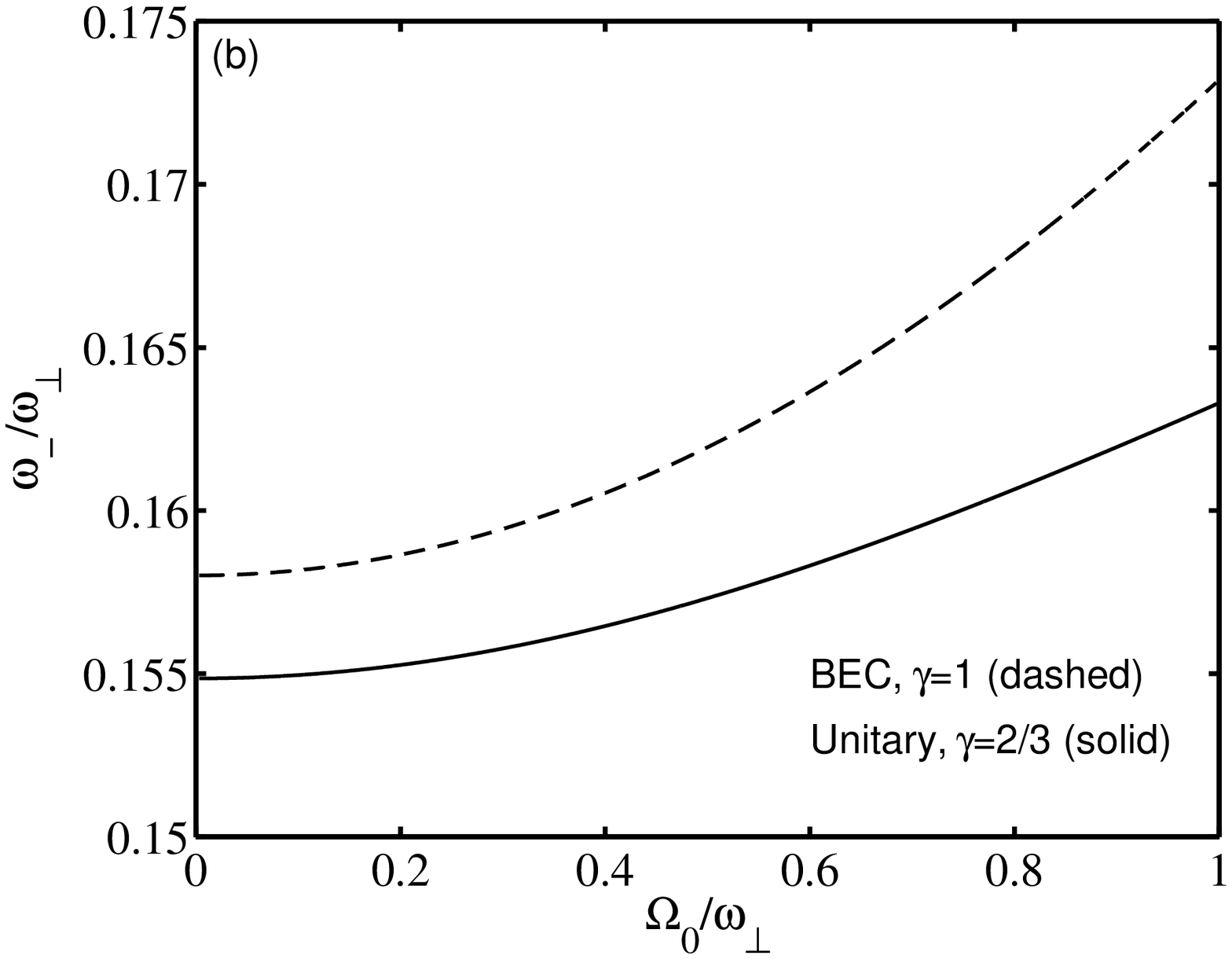}
\caption{\footnotesize Frequencies of the breathing modes $\omega_{+}/\omega_{\perp}$ (a) and $\omega_{-}/\omega_{\perp}$ (b) of Eq.(\ref{finalresult}), as a function of the rotation frequency $\Omega_0/\omega_{\perp}$. We consider a cigar-shape trapping ($\omega_{\perp}/\omega_z=10$), and show results for the BEC ($\gamma=1$) and the Unitary ($\gamma=2/3$)  regimes.}\label{fig:freq}
\end{center}
\end{figure}

The predicted behavior of the frequencies for a typical cigar-shaped geometry
($\omega_\perp/\omega_z=10$) is shown in Fig.\ref{fig:freq}, where 
we explicitly compare the BEC ($\gamma=1$) and the unitary
($\gamma=2/3$) regimes. 

Let us now discuss some  limiting cases predicted by Eq.~(\ref{finalresult}). 
A first important case is given by the centrifugal
limit $\Omega_0\to \omega_\perp$ where the cloud assumes a disk shape as a consequence of the bulge effect  (\ref{bulge}) and  the two solutions (\ref{finalresult}) take the simple form
\begin{eqnarray}
\omega_{+}&=&2\;\omega_{\perp} \ ,
\label{omegazcents}\\
\omega_{-}&=&\sqrt{2+\gamma}\;\omega_{z} \ .
\label{omegazcentsssss}
\end{eqnarray}
In the centrifugal limit the frequency of the radial
breathing mode approaches the universal value $2\omega_\perp$, independent of
the equation of state [see Fig.(\ref{fig:freq})] and of the value of trap deformation
$\omega_\perp/\omega_z$, while the $\gamma$ dependent  frequency of the axial
breathing mode coincides with the value predicted by the hydrodynamic equations
at $\Omega_0=0$, in the disk-shaped configuration $\omega_\perp \ll \omega_z$.
In this regard, one should notice that result (\ref{omegazcentsssss}), as well
as the more general result (\ref{finalresult}), has been derived assuming a 3D
configuration, i.e., assuming the validity of the local density approximation 
along the three directions. When $\Omega_0$ becomes too close to $\omega_\perp$
the gas becomes extremely dilute and the Thomas-Fermi condition $\mu_0\gg
\hbar\omega_z$  is eventually violated with the consequent transition to a 2D
configuration. In this case  the axial frequency takes the ideal gas value
$2\omega_z$ instead of $\sqrt{2+\gamma}\,\omega_z$. For a Bose-Einstein condensed
atomic gas ($\gamma=1$) this transition has been
investigated experimentally \cite{jilacentrfifugal}.
In the case of a Fermi gas at unitarity the conditions required to reach the 2D regime by approaching the centrifugal limit are much more severe (see discussion below).  
  
Another interesting configuration is given by the isotropic trap geometry
$\omega_z=\omega_{\perp}\equiv\omega_0$ for which
Eq.~(\ref{finalresult}) reduces to
\begin{eqnarray}
\lefteqn{\!\!\!\!\!
\omega^2_{\pm} = \frac{4+3\gamma}{2}\;\omega_0^2+(1-\gamma)\;\Omega_0^2}
\nonumber \\
&&\pm\sqrt{\frac{9}{4}\gamma^2\;\omega_0^4+(1-\gamma)^2\;\Omega_0^4
-\gamma(3\gamma-1)\;\omega_0^2\Omega_0^2} \ . 
\label{omegazzeroISO}
\end{eqnarray}
%
%
%
It is worth  noticing that, while in the absence of rotation the frequency
$\omega_-$ reduces to the result $\sqrt2\omega_0$ for the surface quadrupole
$m=0$ mode and the frequency $\omega_+$ approaches the value
$\sqrt{2+3\gamma}\omega_0$ of the pure monopole compression mode, the rotation
provides a coupling between the two modes even for isotropic trapping, so that
they are both sensitive to the value of the equation of state. A special case is
the unitary regime ($\gamma=2/3$) where the two solutions
reduce to
\begin{eqnarray}
\omega_{+}&=&2\;\omega_0 \ ,
\label{omegdded}\\
\omega_{-}&=&\sqrt{2\;\omega_0^2+\frac{2}{3}\;\Omega_0^2} \ .
\label{omegdderrded}
\end{eqnarray}
The spherical trapping geometry in the unitary regime is actually of particular
interest since in this case the Schr\"odinger equation exhibits important
 scaling properties \cite{Werner06}. These give rise to universal
features for the free expansion as well as for the radial monopole frequency
$\omega_+$, which turns out to be independent of $\Omega_0$.
It is worth stressing that, because of the centrifugal effect, the shape of the
gas is not spherical in spite of the spherical symmetry of the trap.
The dynamics are however isotropic, the solution of the equations of motion
being exactly fixed by an isotropic scaling transformation.

Let us now study the experimentally relevant case of a strongly anisotropic
trap $\omega_z\ll\omega_{\perp}$ (cigar shape). In this case
Eq.~(\ref{finalresult}) yields the useful results \cite{notaref}
\begin{eqnarray}
\omega_{+}&=&\sqrt{2(1+\gamma)\;\omega_{\perp}^2+2(1-\gamma)\;\Omega_0^2} \ ,
\label{omegazzero}\\
\omega_{-}&=&
\sqrt{\frac{2+3\gamma+(2-\gamma)\;(\Omega_0/\omega_{\perp})^2}{1+\gamma+(1-\gamma)\;(\Omega_0/\omega_{\perp})^2}}\;\omega_z
\ , \label{omegazzerosse}
\end{eqnarray}
which correspond to the solution for the radial and axial breathing modes,
respectively.
%
%
It is remarkable to see that in the BEC case ($\gamma=1$) the frequency of the
radial breathing mode is independent of $\Omega_0$, reflecting the peculiar
behavior exhibited by the Gross-Pitaevskii equation in 2D \cite{lev}. Conversely,
in the unitary regime ($\gamma=2/3$), the previous equations reduce to
\begin{eqnarray}
\omega_{+}&=&\sqrt{\frac{10}{3}\;\omega_{\perp}^2+\frac{2}{3}\;\Omega_0^2} \ ,
\label{omesejjjjddse}\\
\omega_{-}&=&
2\sqrt{\frac{3+(\Omega_0/\omega_{\perp})^2}{5+(\Omega_0/\omega_{\perp})^2}}\;\omega_z
\ , \label{omeseddspeesise}
\end{eqnarray}
showing that the radial breathing mode depends on the actual value of the
angular velocity and its frequency ranges from the value
$\sqrt{10/3}\omega_\perp$ at $\Omega_0=0$ (recently measured in Lithium gases
\cite{thomas,grimm1,grimm2}) to the universal value $2\omega_\perp$ holding in the centrifugal
limit [see Fig.(\ref{fig:freq})].

It is useful to  recall that the  
hydrodynamic description employed in the present work is based on the applicability
of the Thomas-Fermi  approximation  $\mu_0\gg \hbar\omega_\perp$ which,   in the presence of vortices implies  that the vortex size, fixed by the healing length $\xi$, be much smaller than the inter-vortex distance $d_v=\sqrt{\hbar/M\Omega}$:
\begin{equation}
\xi \ll d_v.
\label{xid}
\end{equation}
%
When $\xi \simeq d_v$ one enters the lowest Landau level (LLL) regime, a regime already explored in rotating Bose gases both theoretically \cite{LLL} and experimentally \cite{jilacentrfifugal}. The condition $\mu_0\gg \hbar\omega_z$ instead ensures the 3D nature of the configuration \cite{note}.  At unitarity, where the healing length is of the order of the interparticle distance $d$,  the transition to the LLL regime ($\xi \simeq d_v$) is reached for  angular velocities extremely close to the centrifugal limit, satisfying the condition 
\begin{equation}
\left[1-\left(\frac{\Omega}{\omega_{0}}\right)^2\right] \simeq \frac{1}{N}.
\label{1/N}
\end{equation}
Here, for simplicity, we have considered isotropic trapping ($\omega_\perp \simeq \omega_z \equiv \omega_{0}$), so that the condition (\ref{1/N}) also implies the transition to 2D.
It is immediate to see that in a Fermi gas at unitarity this condition is equivalent  to requiring that the number of vortices be close to the number of particles  ($N_v\simeq N$) \cite{note:N_v}, a regime where one expects to observe quantum Hall (QH) like effects. 

It is worth comparing Eq.(\ref{1/N}) with the  condition required to reach the LLL and QH regimes in a dilute Bose gas. In this case the  LLL regime (also corresponding to the 2D regime) is obtained under  the less severe requirement $[1-(\Omega/\omega_{0})^2]\simeq 1/G$,  where $G=Na/a_{ho}$. Indeed, since  the scattering length $a$ is much smaller than the oscillator length $a_{ho}=\sqrt{\hbar/M\omega_{0}}$ one has $G\ll N$. The condition for reaching the quantum Hall regime ($N_v\simeq N$) is instead given by the much more severe requirement  $[1-(\Omega/\omega_{0})^2]\simeq G/N^2$. 

An other interesting case concerns the BCS regime of negative values of the
scattering length. As recently pointed out in Ref.~\cite{ho}, in this case the
centrifugal limit cannot be reached by keeping the system in the superfluid
phase. In fact, since the density of the gas becomes smaller and smaller as
$\Omega\to\omega_\perp$, the pairing gap eventually becomes of the order of
the trapping frequency and superfluidity is lost, with the emergence of a
smooth transition between a superfluid central core containing a vortex lattice
and a rotating normal fluid at the periphery. While the equations of rotational
hydrodynamics are expected to hold also in the normal phase, the detailed
structure of elementary excitations might be influenced by the co-existence of
the normal and superfluid components.

Let us finally point out that the equations of hydrodynamics are well suited also to study the problem of the expansion and in particular to predict how the presence of vorticity affects the time evolution of the aspect ratio after release of the trap.

In conclusion we have derived the frequency spectrum of the breathing modes of a 3D harmonically trapped Fermi superfluid in the presence of a vortex lattice. Special attention has been devoted to the unitarity regime, where the collective frequencies are found to exhibit a different dependence on the angular velocity with respect to the case of a dilute Bose gas. 

{\it Acknowledgments}: We gratefully acknowledge stimulating discussions with Lev P. Pitaevskii. We also acknowledge supports by the Ministero dell'Istruzione, dell'Universit\`a e della Ricerca (MIUR).
%


\begin{thebibliography}{99}
\bibitem{notaref}
Equation (\ref{omegazzero}) disagrees with the results of refs. \cite{Machida06} and \cite{Bigelow03}, based on the  irrotationality assumption for the velocity field. In these papers two radial modes with frequencies 
$\sqrt{2(1+\gamma)\left(\omega_{\perp}^2-\Omega_0^2\right)+\Omega_0^2}\pm\Omega_0$
were found. The inclusion of a rotational component in the velocity field [see Eq.(\ref{vANS})] is crucial to ensure that the lowest mode correctly occur at zero frequency, consistently with the conservation of angular momentum.
\bibitem{note} 
In the opposite 2D regime $\mu \ll \hbar \omega_z$ the equation of state of the Fermi gas  exhibits, at unitarity, a linear dependence on the density ($\gamma=1$). In this case the frequency of the radial breathing mode is equal to $2\omega_\perp$, independent of the value of  $\Omega$ while the axial frequency is  equal to $2\omega_z$.
\bibitem{note:N_v}
This can be readily seen by noting that in 2D one has 
${d}^2\simeq R_\perp^2/N\simeq d_{\text{v}}^2N_{\text{v}}/N$ and that, at unitarity, $\xi\simeq d$.
\end{thebibliography}
\end{document}